\documentclass[conference]{IEEEtran}
\usepackage{fancyhdr}
\usepackage{balance}
\usepackage{amsmath}
\usepackage{color}
\usepackage{soul}
\usepackage{mathrsfs}
\usepackage{amssymb}
\usepackage{graphicx}
\usepackage{amsfonts}
\usepackage{booktabs}
\usepackage{algorithm} %format of the algorithm
\usepackage{algorithmic} %format of the algorithm

\DeclareMathOperator*{\argmax}{arg\,max}

\begin{document}
%
% paper title
% can use linebreaks \\ within to get better formatting as desired
% \title{Robust Channel Estimation Method for TDS-OFDM Systems}
\title{Tracking A Dynamic Sparse Channel Via Differential Orthogonal Matching Pursuit}
% author names and affiliations
% use a multiple column layout for up to three different
% affiliations
\author{\IEEEauthorblockN{Xudong Zhu${}^{1}$, Linglong Dai${}^{1}$, Wei Dai${}^{2}$, Zhaocheng Wang${}^{1}$, and Marc Moonen${}^{3}$}
\IEEEauthorblockA{
${}^{1}$Tsinghua National Laboratory for Information Science and Technology (TNlist), Tsinghua University, China\\
%${}^{1}$Department of Electronic Engineering, Tsinghua University, Beijing 100084, China\\
${}^{2}$Department of Electrical and Electronic Engineering, Imperial College, UK\\
${}^{3}$ Department of Electronic Engineering, Katholieke University Leuven, Belgium\\
Email: zhuxd12@mails.tsinghua.edu.cn, wei.dai1@imperial.ac.uk, marc.moonen@esat.kuleuven.be}}

%978-1-4799-0959-9/14/$31.00 ?2014 IEEE
\IEEEoverridecommandlockouts

%\IEEEpubid{\makebox[\columnwidth]{978-1-4799-0959-9/14/\$31.00 ~\copyright~2014 IEEE \hfill}\hspace{\columnsep}\makebox[\columnwidth]{ }}
%\vspace{-0.5cm}

\maketitle

\begin{abstract}
This paper considers the problem of tracking a dynamic sparse channel in a broadband wireless communication system.
A probabilistic signal model is firstly proposed to describe the special features of temporal correlations of dynamic sparse channels: path delays change slowly over time, while path gains evolve faster. %smoothly over time.
Based on such temporal correlations, we then propose the differential orthogonal matching pursuit (D-OMP) algorithm to track a dynamic sparse channel in a sequential way by updating the small channel variation over time.
%When several channel taps suddenly appear or disappear, the proposed D-OMP algorithm is able to converge to the dynamic channel quickly.
Compared with other channel tracking algorithms, simulation results demonstrate that the proposed D-OMP algorithm can track dynamic sparse channels faster with improved accuracy.
%\emph{KeyWords}---Simultaneous multi-channel reconstruction; temporal correlations; compressive sensing (CS); .
\end{abstract}

\IEEEpeerreviewmaketitle

\section{Introduction}
In a broadband wireless communication system, channel state information (CSI) is required at the receiver due to the fact that the multipath fading channel distorts the received signals, especially when the channel is dynamically changing. Hence, accurate channel estimation and tracking become an important problem for communication over a dynamic wireless channel \cite{CEhistory}.

Various linear channel estimation methods with low computational complexity have been proposed \cite{lce}, but their performance is often not robust enough to meet the requirement of communication systems with high rate and high mobility.
Recently, a lot of physical channel measurements have verified that wireless channels exhibit sparsity, i.e., the dimension of a wireless channel may be large, but the number of active taps with significant power is usually small, especially in a broadband wireless communication system \cite{sparse1}.
By exploiting this channel sparsity, many nonlinear channel estimation methods based on classical compressive sensing (CS) algorithms have been proposed to improve the estimation performance, such as orthogonal matching pursuit (OMP), compressive sampling matching pursuit (CoSaMP), and subspace pursuit (SP) \cite{CSCE}-\cite{GG2}.
Compared with the conventional linear methods, CS-based channel estimation methods are able to achieve improved accuracy with reduced training resources \cite{CSCE}.
Further studies have uncovered additional channel characteristics, e.g., the temporal correlations of practical wireless channels: path delays have been shown to change slowly over time, while path gains evolve faster \cite{dynamic2}. %smoothly over time \cite{dynamic2}.
However, the CS-based channel estimation methods ignore these temporal correlations of dynamic sparse channels and have to estimate them independently.
By taking the temporal correlation into account, the adaptive simultaneous OMP (A-SOMP) algorithm has been proposed in \cite{A-SOMP} to obtain simultaneous multi-channel estimates based on the assumption that the dynamic channel estimates in several consecutive time slots share the same path delay set.
However, the path delays of a dynamic sparse channel will change over time, or even there maybe some mutations, so the assumption in \cite{A-SOMP} is not always true in practice.
Another attractive solution for tracking a dynamic sparse channel is the hierarchical Bayesian Kalman (HB-Kalman) filter based on Bayesian CS (BCS) \cite{HB-Kalman}, whereby the iterative re-estimation of the posterior covariance \cite{BayesianCS} is used to achieve accurate channel estimation when sudden changes happen to the dynamic sparse channel, but it suffers from slow tracking speed and high computational complexity.
%Note that compared with the classical CS methods like OMP, CoSaMP, and SP, the A-SOMP algorithm and HB-Kalman algorithm are able to achieve a minor performance gain benefited from the temporal correlations of practical wireless channels.

%In this paper, inspired by the principle of differential encoding \cite{videocoding},
In this paper, we propose a dynamic CS algorithm called differential orthogonal matching pursuit (D-OMP) based on the standard OMP algorithm to track a dynamic sparse channel with fast tracking speed and low computational complexity.
%Differential encoding has been successfully applied to video coding \cite{videocoding}.
%By encoding only the difference between the adjacent frames, redundant information can be dropped. By applying the concept of differential encoding to the standard OMP algorithm, the proposed D-OMP algorithm only needs to detect the small variation of dynamic sparse channels \hl{in a sequential way.}
By exploiting the temporal correlation of a dynamic sparse channel, the proposed D-OMP algorithm only needs to detect the small variation of the dynamic sparse channel in a sequential way.
Furthermore, an adaptive threshold based on the statistical analysis of the ``equivalent" noise is proposed to accurately distinguish true, i.e., non-zero channel taps with low power from thermal noise.
This is essentially different from the standard OMP algorithm in which the incorrect channel taps chosen in one iteration will never be removed, which finally leads to performance degradation.
The performance analysis indicates that a non-zero channel tap can be detected with a high probability, while the noise in the estimate can be removed almost completely. Numerical simulations show that the proposed D-OMP algorithm can track dynamic channels faster and achieve more accurate channel estimates than other tracking algorithms.

The remainder of this paper is organized as follows. The system model of dynamic sparse channels is described in Section II. Section III addresses the proposed D-OMP algorithm, together with the threshold based on noise statistics. Section IV presents the performance analysis, and simulation results are provided in Section V. Finally, conclusions are drawn in Section VI.

\emph{Notation}:
We use upper-case and lower-case boldface letters to denote matrices and vectors, respectively;
$(\cdot)^T$, $(\cdot)^H$, $(\cdot)^{-1}$, $(\cdot)^{\dagger}$, and $\|\cdot\|_p$ denote the transpose, conjugate transpose, matrix inversion, Moore-Penrose matrix inversion, and $l_p$ norm operation, respectively;
$|\Gamma|$ denotes the number of elements in set $\Gamma$;
$\mathbf{h}_{\Gamma}$ denotes the entries of the vector $\mathbf{h}$ in the set $\Gamma$;
$\mathbf{\Phi}_{\Gamma}$ denotes the submatrix comprising the $\Gamma$ columns of $\mathbf{\Phi}$.

\section{System Model}
\begin{figure}
\vspace{-2.5mm}
\center{\includegraphics[width=0.5\textwidth]{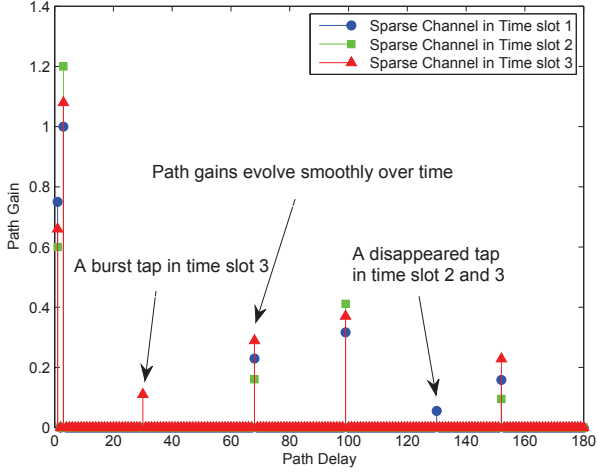}}
\vspace{-3mm}
\caption{Illustration of the dynamic Vehicular B channel with a velocity of 120 km/h in three consecutive time slots.}
\vspace{-3mm}
\label{Dynamic_Sparse_Channel}
\end{figure}

In this paper, we consider the problem of tracking a dynamic sparse channel $\{\mathbf{h}^{(t)}\}_{t=1}^T$ from under-sampled noisy measurements $\{\mathbf{y}^{(t)}\}_{t=1}^T$ in $T$ time slots, in a broadband wireless communication system. The single time slot measurement vector $\mathbf{y}^{(t)}=[y_1^{(t)}, y_2^{(t)}, \cdots, y_M^{(t)}]^T$ is usually obtained through the linear measurement process
\begin{equation}
\label{y=phih+n}
\mathbf{y}^{(t)}=\mathbf{\Phi}^{(t)}\mathbf{h}^{(t)}+\mathbf{n}^{(t)},\hspace{0.5cm} t=1, 2, \cdots, T,
\end{equation}
where $\mathbf{h}^{(t)}=[h_1^{(t)}, h_2^{(t)}, \cdots, h_N^{(t)}]^T$ denotes the time-domain discrete channel vector with $N$ taps, where $N>M$, and the additive white Gaussian noise (AWGN) is modeled as $\mathbf{n}^{(t)}\sim\mathcal{CN}(\mathbf{0},\sigma_n^2 \mathbf{I}_M)$, where $\mathbf{I}_M$ denotes the identity matrix of size $M\times M$.
The measurement matrix $\mathbf{\Phi}^{(t)}$ of size $M\times N$ is a Toeplitz matrix determined by the training sequence $\mathbf{c}=[c_0,c_1, \cdots, c_M]^T$ in the time or frequency domain, e.g., the well known pseudo noise (PN) sequence used in CDMA systems \cite{CSCE}.
Thus the measurement matrix $\mathbf{\Phi}^{(t)}$ becomes time-invariant \cite{A-SOMP}, i.e., $\mathbf{\Phi}^{(1)}=\mathbf{\Phi}^{(2)}=\cdots=\mathbf{\Phi}^{(T)}=\mathbf{\Phi}=[\boldsymbol{\phi}_1,\boldsymbol{\phi}_2, \cdots, \boldsymbol{\phi_N}]$.
Since the training sequence is usually designed in a random way, i.e., $c_i\sim\mathcal{CN}(0, \sigma_c^2)$, we can obtain that $\phi_{i,j}^{(t)}\sim\mathcal{CN}(0, \sigma_{\phi}^2)$, where $\sigma_{\phi}^2=\sigma_c^2$ and $\phi_{i,j}^{(t)}$ denotes the element of the matrix $\mathbf{\Phi}^{(t)}$.
%Once the training sequence is determined, the measurement matrix $\mathbf{\Phi}^{(t)}$ becomes time-invariant, i.e., $\mathbf{\Phi}^{(1)}=\mathbf{\Phi}^{(2)}=\cdots=\mathbf{\Phi}^{(T)}=\mathbf{\Phi}=[\phi_1,\phi_2, \cdots, \phi_N]$.
%Furthermore, the columns of $\mathbf{\Phi}$ are assumed to satisfy a semi-orthogonality, i.e., $\boldsymbol{\phi}_i^H\boldsymbol{\phi}_j\approx 0, i\neq j$, due to the randomness of the training sequence.
Furthermore, based on this Gaussian assumption, the columns of the $\mathbf{\Phi}$ matrix are semi-orthogonal, i.e., $\boldsymbol{\phi}_i^H\boldsymbol{\phi}_j\approx 0, i\neq j$.
%The measurement matrix $\mathbf{\Phi}^{(t)}$ of size $M\times N$ is determined by the training sequence in the time or frequency domain and is usually time-invariant \cite{CSCE}, \cite{A-SOMP}.
%Hence we assume that $\mathbf{\Phi}^{(1)}=\mathbf{\Phi}^{(2)}=\cdots=\mathbf{\Phi}^{(T)}=\mathbf{\Phi}=[\phi_1,\phi_2, \cdots, \phi_N]=[\phi_{i,j}]_{M\times N}$ is a known measurement matrix.
%\hl{Furthermore, since the training sequence is usually designed in a pseudo random way, e.g., the well known pseudo noise (PN) sequence used in CDMA systems, here we assume the element $\phi_{i,j}$ is independent and identically distributed (i.i.d.), i.e., $\phi_{i,j}\sim\mathcal{CN}(0, \sigma_{\phi}^2)$.}

The temporal correlations of practical wireless channels have been verified through analysis and experiments, even when the channels are varying fast \cite{dynamic2}. Fig. \ref{Dynamic_Sparse_Channel} illustrates the time-domain impulse response of the dynamic Rayleigh fading Vehicular B channel with a velocity of 120 km/h in three consecutive time slots \cite{Vehicular}. It is clear that path delays of such dynamic sparse channels change slowly, while path gains evolve faster. In order to characterize the temporal correlations of a dynamic sparse channel, we adopt a probabilistic signal model with two time series vectors $\{\mathbf{s}^{(t)}\}_{t=1}^{T}$ and $\{\mathbf{a}^{(t)}\}_{t=1}^{T}$, where the binary vector $\mathbf{s}^{(t)}=[s_1^{(t)}, s_2^{(t)}, \cdots, s_N^{(t)}]^T$ is used to describe the temporal evolution of the channel paths, while the complex-valued vector $\mathbf{a}^{(t)}=[a_1^{(t)}, a_2^{(t)}, \cdots, a_N^{(t)}]^T$ characterizes the temporal variation of the path gains. Hence, the dynamic sparse channel can be modelled as
\begin{equation}
\label{h=salpha}
h_i^{(t)}=s_i^{(t)} a_i^{(t)},\hspace{0.5cm} t=1, 2, \cdots, T, \hspace{0.2cm}1\leq i\leq N
\end{equation}
where the binary $s_i^{(t)}\in\{0, 1\}$ denotes whether there is a non-zero channel tap at index $i$ in the $t$th time slot, and $a_i^{(t)}$ denotes the corresponding path gain. Thus, the path delay set $\Lambda^{(t)}$ of the dynamic sparse channel can be represented as $\Lambda^{(t)}=\{i:s_i^{(t)}=1\}$.

We model the time series $\{\mathbf{s}^{(t)}\}_{t=1}^{T}$ as a discrete Markov process with two transition probabilities \cite{Markov}: $p_{0\rightarrow 1}=P\{s_i^{(t)}=1|s_i^{(t-1)}=0\}$ and $p_{1\rightarrow 0}=P\{s_i^{(t)}=0|s_i^{(t-1)}=1\}$. Without loss of generality, $\{\mathbf{s}_i^{(1)}\}_{i=1}^{N}$ is initialized as independent Bernoulli random variables based on the probability $p_1$, i.e., $s_i^{(1)}\sim$Bernouli$(p_1), i=1, 2, \cdots, N$.
%Thus, $p_{0\rightarrow 1}$ and $p_{1\rightarrow 0}$ represent the change rate of path delays, and $p_{1}$ favors how sparse the dynamic sparse channel is.
Similarly, we model the time series $\{\mathbf{a}^{(t)}\}_{t=1}^{T}$ as a Gauss-Markov process, and each path gain evolves independently as
\begin{equation}
\label{a=a+w}
a_i^{(t)}=a_i^{(t-1)}+w_i^{(t)},\hspace{0.5cm} t=1, 2, \cdots, T, \hspace{0.2cm}1\leq i\leq N
\end{equation}
where $\mathbf{w}^{(t)}=[w_1^{(t)}, w_2^{(t)}, \cdots, w_N^{(t)}]^T\sim\mathcal{CN}(\mathbf{0}, \sigma_a^2\mathbf{I}_N)$ controls the temporal correlation of the path gains \cite{Markov}. We model the initial distribution of path gains as a Gaussian distribution, i.e., $\{a_i^{(1)}\}_{i=1}^{N}\sim\mathcal{CN}(\mathbf{0}, \sigma_h^2\mathbf{I}_N)$. %The above mentioned parameters will be set in detail in the simulation section.

The probabilities $p_{0\rightarrow 1}$ and $p_{1\rightarrow 0}$ are usually small to model the slow changing of the path delays, while $\{\mathbf{a}^{(t)}\}_{t=1}^T$ changes in each time slot to model the faster changing of the path gains. Furthermore, the aforementioned parameters will be given in detail in Section V.

\section{Proposed D-OMP Algorithm}
In this section, the proposed D-OMP algorithm is explained in detail. Then the threshold used in the algorithm is derived based on noise statistics.

\subsection{D-OMP Algorithm}
%Inspired by the concept of differential encoding,
We propose the D-OMP algorithm to track a dynamic sparse channel, which obtains the final estimates in a sequential way by updating the small variation of the dynamic sparse channel.
The key idea of the proposed D-OMP algorithm is that, the major information of the channel in the current time slot can be obtained from the estimation results in previous time slots due to the temporal correlations of the dynamic sparse channel, and then the small variation of the channel can be estimated with low complexity to refine the final estimate result.
This is essentially different from the standard OMP algorithm which ignores any temporal correlation of the channel taps \cite{StandardOMP}.
The proposed algorithm also differs from the HB-Kalman algorithm utilizing the iterative re-estimation of the posterior covariance to obtain channel estimation, which leads to a slow tracking speed and high computational complexity.

\begin{algorithm}[htb]
\renewcommand{\algorithmicrequire}{\textbf{Input:}}
\renewcommand\algorithmicensure {\textbf{Output:} }
\caption{D-OMP Algorithm}
\label{D-OMP}
\begin{algorithmic}[1]
\REQUIRE ~~\\
Received signals: $\{\mathbf{y}^{(t)}\}_{t=1}^{T}$;\\
Measurement matrix: $\mathbf{\Phi}=[\boldsymbol{\phi}_{1}, \boldsymbol{\phi}_{2}, \cdots, \boldsymbol{\phi}_{N}]$; \\
Threshold: $P_{th}$.\\
\ENSURE ~~\\
Channel estimates: $\{\mathbf{\hat{h}}^{(t)}\}_{t=1}^{T}$. \\
\STATE Initialization : \\
\STATE $\;\;\;\;\hat{\Lambda}^{(0)}=\varnothing$, $\hat{\mathbf{h}}^{(0)}=\mathbf{0}$.
\FOR {$t=1$ to $T$}
\STATE $\mathbf{y}^{+}=\mathbf{y}^{(t)}-\mathbf{\Phi}\hat{\mathbf{h}}^{(t-1)}$. %\hspace{4.9cm}%\% incremental observations
\STATE $\Lambda^{+}=\argmax_{i} \{\lambda_i=|\boldsymbol{\phi}_{i}^H \mathbf{y}^{+}|: i\notin \hat{\Lambda}^{(t-1)}\}$. %\hspace{0.95cm} %\% incremental support detection
\STATE $\hat{\Lambda}^{(t)}=\hat{\Lambda}^{(t-1)}\cup\Lambda^{+}$.%\hspace{5.2cm} %\% temporary path delays set
\STATE $\hat{\mathbf{h}}^{(t)}=\arg\min_{\mathbf{h}}\{\|\mathbf{\Phi}\mathbf{h}-\mathbf{y}^{(t)}\|_2: supp(\mathbf{h})\subseteq\hat{\Lambda}^{(t)}\}$.  %\% channel estimate update
\STATE $\Lambda^{-}=\{i:|\hat{h}_i^{(t)}|\leq P_{th}, i\in\hat{\Lambda}^{(t)}\} $. %\hspace{2.85cm} %\% disappearing channel taps detection
\STATE $\hat{\mathbf{h}}_{\Lambda^{-}}^{(t)}=\mathbf{0}$, $\hat{\Lambda}^{(t)}=\hat{\Lambda}^{(t)}-\Lambda^{-}$. %\hspace{3.65cm} %\% disappearing channel taps removal
\ENDFOR
\RETURN $\{\mathbf{\hat{h}}^{(t)}\}_{t=1}^{T}$.
\end{algorithmic}
\end{algorithm}

The pseudocode of the proposed D-OMP algorithm is provided in Algorithm 1, and each loop is comprised of the following three parts:

\subsubsection{Incremental Support Detection (step 4$\sim$6)}
Firstly, we obtain the incremental observations $\mathbf{y}^{+}$ by subtracting $\mathbf{\Phi}\hat{\mathbf{h}}^{(t-1)}$ from $\mathbf{y}^{(t)}$ in step 4, where $\hat{\mathbf{h}}^{(t-1)}$ denotes the estimate of $\mathbf{h}^{(t-1)}$.
Here we use $\mathbf{\Phi}\hat{\mathbf{h}}^{(t-1)}$ instead of $\mathbf{y}^{(t-1)}$ to avoid the impact of the noise $\mathbf{n}^{(t-1)}$.
Different from the way used in the standard OMP algorithm which iteratively detects the candidate supports by selecting the element that correlates best with the residual signal \cite{StandardOMP}, here we only detect the incremental support $\Lambda^{+}$ in step 5, which is able to acquire the appearing channel tap as $s_i^{(t-1)}=0\rightarrow s_i^{(t)}=1$.
Then, the temporary estimate of the path delays set $\hat{\Lambda}^{(t)}$ is obtained by merging $\hat{\Lambda}^{(t-1)}$ which contains the persistent channel taps, i.e., $i\in\Lambda^{(t-1)}\cap\Lambda^{(t)}$, and $\Lambda^{+}$ which contains the appearing channel tap.

\subsubsection{Channel Estimate Update (step 7)}
Step 7 aims to update the channel estimate based on the temporary estimate of the path delays set $\hat{\Lambda}^{(t)}$.
Intuitively, the variation of the channel estimate is comprised of three parts: a) The appearing channel tap $a_i^{(t)}$ when $s_i^{(t-1)}=0\rightarrow s_i^{(t)}=1$; b) The disappearing channel tap $-a_j^{(t-1)}$ when $s_j^{(t-1)}=1\rightarrow s_j^{(t)}=0$; c) The smooth variation of the path gains $w_i^{(t)}(i\in\hat{\Lambda}^{(t)})$.
The minimization problem in step 7 corresponds to the main computational burden of the D-OMP algorithm, which can be realized by the standard least-squares (LS) technique with low complexity \cite{GS}.
%Then, in step 8, we obtain the temporary channel estimate by merging the previous channel estimate $\hat{\mathbf{h}}^{(t-1)}$ and the channel variation $\mathbf{h}^{+}$.

\subsubsection{Disappearing Channel Taps Removal (step 8$\sim$9)}
When a tap $h_i^{(t-1)}$ disappears as $s_i^{(t-1)}=1\rightarrow s_i^{(t)}=0$, the estimated tap in the previous $(t-1)$th time slot will still remain in the estimated path delay set of the current time slot since $\hat{\Lambda}^{(t)}=\hat{\Lambda}^{(t-1)}\cup\Lambda^{+}$.
 %and this ``fake tap" is difficult to be removed, especially when the disappearing tap usually has low power.
%Fortunately,
The path gain of this disappearing channel tap $\hat{h}_i^{(t)}$ will be small.
 %since it is the sum of $\hat{h}_i^{(t-1)}$ and $h_i^{+}$ which have similar absolute values but opposite signs in the proposed D-OMP algorithm, i.e., $\hat{h}_i^{+}\approx-\hat{h}_i^{(t-1)}$ due to the operation $\mathbf{\Phi}\mathbf{h}-\mathbf{y}^{+}$ in step 7.
Therefore, we propose a threshold $P_{th}$ to judge whether the nonzero elements in $\hat{\mathbf{h}}^{(t)}$ are non-zero channel taps or noise in step 8.
%What's more, the incremental support detection $\Lambda^{+}$ will be a noise when there is no burst taps and it will also be removed by appropriately selecting $P_{th}$.
After removing the disappearing channel taps, the final channel estimate can be obtained in step 9.

Unlike the standard OMP algorithm which needs to solve the minimization problem for many times according to the sparsity level of the wireless channel \cite{GG1}, the proposed D-OMP algorithm only needs to solve the minimization problem once, so the computational complexity can be dramatically reduced. Furthermore, unlike the hard threshold \cite{CSCE} or the rough criterion \cite{A-SOMP}, we rely on the noise analysis to select the threshold $P_{th}$, which will be derived in the following subsection.

\subsection{Threshold Based on Noise Statistics}
Here we consider a single time slot (we omit the time slot superscript $t$), and rewrite (\ref{y=phih+n}) as
\begin{equation}
\label{y=phi(h+n)}
\mathbf{y}=\mathbf{\Phi}\mathbf{h}+\mathbf{n}=\mathbf{\Phi}(\mathbf{h}+\mathbf{\Phi}^{\dag}\mathbf{n})=\mathbf{\Phi}(\mathbf{h}+\mathbf{n}^{'}),
\end{equation}
%\begin{eqnarray}
%\label{y=phi(h+n)}
%\nonumber \mathbf{y}&=&\mathbf{\Phi}\mathbf{h}+\mathbf{n}\\
%\nonumber &\approx& \mathbf{\Phi}(\mathbf{h}+\mathbf{\Phi}^{\dag}\mathbf{n})\\
%&=& \mathbf{\Phi}(\mathbf{h}+\mathbf{n}^{'}),
%\end{eqnarray}
where $\mathbf{\Phi}^{\dag}=\mathbf{\Phi}^H(\mathbf{\Phi}\mathbf{\Phi}^H)^{-1}$ denotes the pseudoinverse of $\mathbf{\Phi}$.
%, and we use the symbol $\approx$ in (4) due to $\mathbf{\Phi}\mathbf{\Phi}^{\dag}$ is not strictly but approximately equal to $\mathbf{I}_M$.

Generally, linear channel estimation methods are designed to directly approach $\mathbf{h}+\mathbf{n}^{'}$, so their performance is limited by the contamination of the ``equivalent" noise $\mathbf{n}^{'}=\mathbf{\Phi}^{\dagger}\mathbf{n}$ in (4).
The advantage of CS-based nonlinear estimation methods is that they reconstruct the sparse channel $\mathbf{h}$ and avoid part of the interference caused by the ``equivalent" noise $\mathbf{n}^{'}$.
Since the basic idea of the greedy CS methods is to iteratively search the channel taps from the strongest one to the weakest one \cite{CSCE}, \cite{A-SOMP}, it is difficult for these algorithms to judge whether the searched taps are true, i.e., non-zero channel taps or noise, when they are weak.

In order to further reduce the interference caused by $\mathbf{n}^{'}$ under the CS framework, we design the threshold $P_{th}$ based on the statistical properties of the noise $\mathbf{n}^{'}$ for the proposed D-OMP algorithm to separate non-zero channel taps from noise as much as possible.
According to (\ref{y=phi(h+n)}), we have $\mathbf{n}^{'}=\mathbf{\Phi}^{\dagger}\mathbf{n}$.
The mean of the vector $\mathbf{n}^{'}$ can be easily obtained as $\text{E}\{\mathbf{n}^{'}\}=\text{E}\{\mathbf{\Phi}^{\dagger}\}\text{E}\{\mathbf{n}\}=\mathbf{0}$ due to the fact that the measurement matrix $\mathbf{\Phi}$ and the noise vector $\mathbf{n}$ are independent.
Then, the covariance of the elements in equivalent noise $\mathbf{n}^{'}$ can be derived as %$\text{Cov}\{\mathbf{n}^{'}\}=\text{Var}\{n_i^{'}\}\mathbf{I}_N$, where $\text{Var}\{n_i^{'}\}$ can be derived as
\begin{eqnarray}
\text{Var}\{n_i^{'}\}&=& \text{Var}\{\boldsymbol{\phi}_{i}^{\dagger}\mathbf{n}\}= M\text{Var}\{\phi_{i,j}^{\dagger}\}\text{Var}\{n_j\}\nonumber\\
&=&M\sigma_n^2\text{Var}\{\phi_{i,j}^{\dagger}\},
\end{eqnarray}
where $\boldsymbol{\phi}_{i}^{\dagger}$ denotes the $i$th row of the matrix $\mathbf{\Phi}^{\dagger}$ and $\phi_{i,j}^{\dagger}$ is the $j$th element of $\boldsymbol{\phi}_{i}^{\dagger}$. Furthermore, we have
\begin{equation}
\mathbf{\Phi}^{\dagger}=\mathbf{\Phi}^H(\mathbf{\Phi}\mathbf{\Phi}^H)^{-1}
\approx\mathbf{\Phi}^H(N\sigma_{\phi}^2\mathbf{I}_M)^{-1}
=\frac{1}{N\sigma_{\phi}^2}\mathbf{\Phi}^H,
\end{equation}
where we use $N\sigma_{\phi}^2\mathbf{I}_M$ to approximate $\mathbf{\Phi}\mathbf{\Phi}^H$ due to the randomness of the matrix $\mathbf{\Phi}$.
Then we can obtain the variance of $n_i^{'}$ as
\begin{equation}
\text{Var}\{n_i^{'}\}\approx\frac{M\sigma_n^2}{(N\sigma_{\phi}^2)^2}\text{Var}\{\phi_{i,j}^{*}\}=\frac{M\sigma_n^2}{N^2\sigma_{\phi}^2}.
\end{equation}
Finally, the statistical properties of the elements in $\mathbf{n}^{'}$ can be represented as
\begin{equation}
\label{n'=n}
n_i^{'}\sim\mathcal{CN}(0, \sigma_{n^{'}}^2)\approx\mathcal{CN}(0,\frac{M\sigma_n^2}{N^2\sigma_{\phi}^2}).
\end{equation}
\iffalse
\begin{equation}
\label{n'=n}
\mathbf{n}^{'}\sim\mathcal{CN}(\mathbf{0}, \sigma_{n^{'}}^2\mathbf{I}_N)\approx\mathcal{CN}(\mathbf{0},\frac{M\sigma_n^2}{N^2\sigma_{\phi}^2}\mathbf{I}_N).
\end{equation}
\fi

With the help of the statistical properties of the noise vector $\mathbf{n}^{'}$, the threshold $P_{th}$ can be selected as
\begin{equation}
\label{th}
P_{th} = \alpha\sigma_{n^{'}}\approx\frac{\alpha\sigma_{n}\sqrt{M}}{\sigma_{\phi}N},
\end{equation}
where the variance $\sigma_n^2$ of the noise vector $\mathbf{n}$ can be usually obtained at the receiver \cite{SNRDetection}.
The coefficient $\alpha$ can be set as $\alpha=3$ so that the strength of the noise element $|n_i^{'}|$ will be smaller than $P_{th}$ with high probability of $99.73\%$, based on the assumed Gaussian distribution.

%In the proposed D-OMP algorithm, the support set $\hat{\Lambda}^{(t)}$ will be contaminated by the noise in two ways: the first one is that the incremental support detection $\Lambda^{+}$ when there is no burst taps; the second one is that the pre-estimate support set $\hat{\Lambda}^{(t-1)}$ when there is a disappeared tap. The threshold $P_{th}$ used in the propose D-OMP algorithm is able \hl{to remove the fake taps which are comparable with the noise $\mathbf{n}^{'}$.}

\section{Performance Analysis}

%\subsection{Correct Detection Probability}
\subsection{Persistent Channel Tap Detection}
In section III-B, the proposed threshold $P_{th}=\alpha\sigma_{n^{'}}$ is used to remove the disappearing channel taps from the channel estimate result with high probability.
On the other hand, as the proposed scheme may also remove a persistent channel tap by mistake, the correct detection probability of the persistent tap for the scheme is also important to guarantee its performance in practice, which will be analyzed below.

Firstly, the SNR $\gamma$ at the receiver can be obtained \cite{SNRDetection} and represented as
\begin{equation}
\label{SNR}
\gamma=10\text{log}\frac{\text{Var}\{\mathbf{\Phi}\mathbf{h}\}}{\text{Var}\{\mathbf{n}\}}=10\text{log}\frac{K\sigma_{\phi}^2\sigma_h^2}{\sigma_n^2}=10\text{log}\frac{M K\sigma_h^2}{N^2\sigma_{n^{'}}^2},
\end{equation}
where $K=N p_{1}$ denotes the sparsity level of the dynamic sparse channel. Thus, the threshold $P_{th}$ can be obtained as
\begin{equation}
P_{th}=\alpha\sigma_{n^{'}}=\alpha\sigma_h\sqrt{\frac{M K}{10^{\frac{\gamma}{10}}N^2}}.
\label{P_th}
\end{equation}

After the threshold has been obtained, we can calculate the detection probability of a persistent channel tap as $P(h_i^{(t)}\in \Lambda)$, which can be derived by using the normal Gaussian distribution probability function $\Psi$ as
\begin{eqnarray}
P(h_i^{(t)}\in \Lambda)&=&P(|h_i^{(t)}|>P_{th})\nonumber\\
&=&2\left(1-\Psi(\alpha\sqrt{\frac{M K}{10^{\frac{\gamma}{10}}N^2}})\right).
\label{theory_error}
\end{eqnarray}
For example, in a typical wireless communication system with $M=200$, $N=400$, $K=5$, and $\gamma=15$ dB, the probability of a persistent tap being larger than $P_{th}$ is about $96.64\%$.

%\subsection{\hl{Condition of Incremental Support Detection}}
\subsection{Appearing Channel Tap Detection}
As the D-OMP algorithm is proposed to track a dynamic channel, the condition of appearing channel tap detection is important to be analyzed.
For the sake of a simplified analysis, we assume that the channel estimate in time slot $(t-1)$ is accurate without loss of generality, i.e., $\mathbf{\hat{h}}^{(t-1)}=\mathbf{h}^{(t-1)}$.
When there is an appearing channel tap from time slot $(t-1)$ to $(t)$ (for simplicity, we assume other channel taps stay unchanged, i.e., $\mathbf{h}_k^{(t)}=\mathbf{h}_k^{(t-1)}, \forall k\in\Lambda^{(t-1)}$), the step 4 in Algorithm 1 can be rewritten as
%\hl{We assume that there is an appearing channel tap} from time slot $(t-1)$ to $(t)$, without loss of generality, and the channel estimate in time slot $(t-1)$ is accurate, i.e., $\mathbf{\hat{h}}^{(t-1)}=\mathbf{h}^{(t-1)}$, for the sake of a simplified analysis. Thus, we can rewrite step 4 in Algorithm 1 as
\begin{equation}
\mathbf{y}^{+}=\mathbf{y}^{(t)}-\mathbf{\Phi}\mathbf{h}^{(t-1)}=\boldsymbol{\phi}_{i}h_i^{(t)}+\mathbf{n},%(\mathbf{n}^{\prime}),
\end{equation}
where $h_i^{(t)}$ ($i\notin\Lambda^{(t-1)}$) denotes the appearing channel tap at the time index $i$ during the $t$th time slot of the dynamic sparse channel.

According to the detection rule in step 5 of Algorithm 1, the appearing channel tap $h_i^{(t)}$ is successfully detected if and only if
\begin{equation}
\label{igeqj}
\lambda_j<\lambda_i,\hspace{0.3cm} \forall j\neq i, j\notin\hat{\Lambda}^{(t-1)},
\end{equation}
where $\lambda_i$ can be derived as
\begin{equation}
\label{lambda_i}
\lambda_i=|\boldsymbol{\phi}_i^{H}\mathbf{y}^{+}|=|\boldsymbol{\phi}_i^{H}(\boldsymbol{\phi}_i h_i^{(t)}+\mathbf{n})|\geq|\boldsymbol{\phi}_i^{H}\boldsymbol{\phi}_i| |h_i^{(t)}|-|\boldsymbol{\phi}_i^{H}\mathbf{n}|.
%\overset{\beta}{\geq} \sigma_{\phi}^2M\|h_i^{(t)}\|_2-\beta\sigma_{\phi}\sigma_n\sqrt{M},
\end{equation}
%where the triangle inequality principle is applied here.
%and the symbol $\beta$ means that the establishment probability of the inequality is based on $\beta$. Similar with that in equation (\ref{th}), the probability is $99.73\%$ if we take $\beta=3$.
Then $\lambda_j$ can be similarly derived as
\begin{equation}
\label{lambda_j}
\lambda_j=|\boldsymbol{\phi}_j^{H}\mathbf{y}^{+}|=|\boldsymbol{\phi}_j^{H}(\boldsymbol{\phi}_i h_i^{(t)}+\mathbf{n})|\leq |\boldsymbol{\phi}_j^{H}\boldsymbol{\phi}_i| |h_i^{(t)}|+|\boldsymbol{\phi}_j^{H}\mathbf{n}|.
%\overset{\beta}{\leq}\beta\sigma_{\phi}^2\sqrt{M}\|h_i^{(t)}\|_2+\beta\sigma_{\phi}\sigma_n\sqrt{M},
\end{equation}
Using (\ref{igeqj}), we can obtain the condition under which the appearing channel tap $h_i^{(t)}$ is successfully detected:
%with a high probability based on $\beta$ when
\begin{equation}
\label{h_condition}
|h_i^{(t)}|\geq\frac{|\boldsymbol{\phi}_i^{H}\mathbf{n}|+|\boldsymbol{\phi}_j^{H}\mathbf{n}|}{|\boldsymbol{\phi}_i^{H}\boldsymbol{\phi}_i|-|\boldsymbol{\phi}_j^{H}\boldsymbol{\phi}_i|}, \hspace{0.5cm} \forall j\neq i, j\notin\hat{\Lambda}^{(t-1)},
\end{equation}
which means the path gain of the appearing channel tap should be larger than a minimum value.
 %to be correctly detected from the noisy measurements.
%\hl{whether to add some context to explain the equation (17).}

Intuitively, the idea behind the incremental support detection of the proposed D-OMP algorithm is to find the elements $\Lambda^{+}$ that correlate best with the incremental observations $\mathbf{y}^{+}$, but the previously detected support $\hat{\Lambda}^{(t-1)}$ will not be considered.
Due to the fact that the appearing channel tap can be detected immediately if it is larger than a certain value derived in (\ref{h_condition}), while the posterior covariance matrix of the HB-Kalman method \cite{HB-Kalman} needs several time slots to converge, the condition of incremental support detection of the proposed D-OMP algorithm is much easier to satisfy than that of the HB-Kalman method.

\subsection{Computational Complexity}
The computational complexity of the proposed D-OMP algorithm in terms of the required number of complex multiplications includes the following three parts:
\begin{enumerate}
  \item In the incremental detection part, the complexity is $\mathcal{O}(NM)$ for the calculation of $\mathbf{\Phi}\hat{\mathbf{h}}^{(t-1)}$ and $|\boldsymbol{\phi}_i^H\mathbf{y}^{+}|$.
  \item In the channel estimate update part, the LS problem $\arg\min_{\mathbf{h}}\{\|\mathbf{\Phi}\mathbf{h}-\mathbf{y}^{+}\|_2: supp(\mathbf{h})\subseteq\hat{\Lambda}^{(t)}\}$ can be implemented with the $\mathcal{O}(MK^2)$ complexity by using the Gram-Schmidt algorithm \cite{GS}.
  \item In the disappearing taps removal part, the complexity is $\mathcal{O}(M)$ for the comparison of taps with the threshold $P_{th}$.
\end{enumerate}

To sum up, the total complexity of the proposed D-OMP algorithm is $\mathcal{O}(M(N+K^2+1)T)$ to track a dynamic sparse channel over $T$ time slots.
Compared with the $\mathcal{O}(KM(N+K^2)T)$ complexity of the standard OMP algorithm, the complexity of the proposed D-OMP algorithm is reduced approximately by a factor of $K$.
Considering the re-estimation of the posterior covariance matrix with $J$ iterations in each time slot \cite{BayesianCS}, the computational complexity of HB-Kalman filter \cite{HB-Kalman} is $\mathcal{O}(JKM(N+K^2)T)$, which is much higher than the complexity for both the standard OMP algorithm and the proposed D-OMP algorithm.
The complexity reduction comes from the fact that the proposed D-OMP algorithm obtains the major information of the current channel from the estimation results in previous time slots, and then refines the final estimation result by detecting the small variation of the channel.

\section{Simulation Results and Discussion}
This section investigates the performance of the proposed D-OMP algorithm compared with the conventional linear algorithm \cite{lce}, standard OMP algorithm \cite{GG1}, A-SOMP algorithm \cite{A-SOMP}, and HB-Kalman algorithm \cite{HB-Kalman}.
The parameters mentioned in Section II (system model) are set as follows: 1) The size of the matrix $\mathbf{\Phi}$ is $N=400$ and $M=200$; 2) The probability $p_1$ is set as 0.025, which means the average channel sparsity level is $K=N p_1=10$; 3) A channel tap will appear or disappear over every $1/(K p_{1\rightarrow 0})=1/((N-K) p_{0\rightarrow 1})=10$ time slots on average corresponding to probabilities $p_{1\rightarrow0}=0.01$ and $p_{0\rightarrow1}=p_1p_{1\rightarrow0}/(1-p_1)$; 4) $\sigma_a=0.05, \sigma_n=0.05, \sigma_h=1, \sigma_{\phi}=1$.

%The parameters mentioned in Section II (system model) are set as: 1) $N=400, M=200$; 2) $p_1=0.025, p_{1\rightarrow0}=0.01, p_{0\rightarrow1}=p_1p_{1\rightarrow0}/(1-p_1)$; 3) $\sigma_a=0.05, \sigma_n=0.05, \sigma_h=1, \sigma_{\phi}=1$. Thus, the average channel sparsity level is $K=Np_1=10$. A tap will appear or disappear over every 10 time slots on average due to the probabilities $p_{1\rightarrow0}$ and $p_{0\rightarrow1}$.

\begin{figure}
\center{\includegraphics[width=0.5\textwidth]{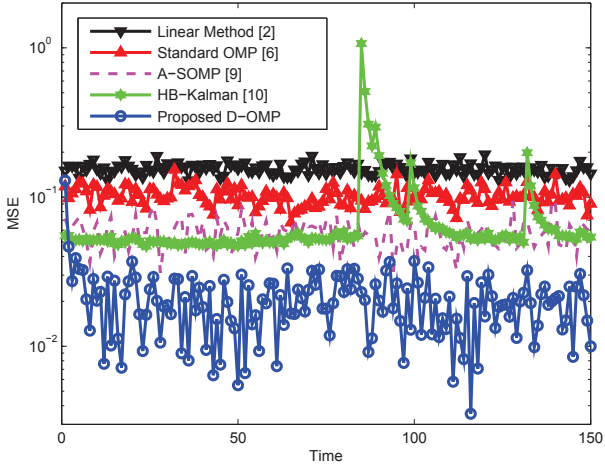}}
\vspace{-5mm}
\caption{MSE performance comparison against time slot $t$.}
\vspace{-3mm}
\label{MSE_Time_full}
\end{figure}

Fig. \ref{MSE_Time_full} shows the mean squared error (MSE) performance against time slot $t$ for the five channel estimation methods mentioned above. It is clear that A-SOMP and HB-Kalmam achieve lower error levels than the standard OMP algorithm, while the conventional linear method performs worst.
The MSE performance of the proposed D-OMP algorithm is the best, as the temporal correlations of the dynamic sparse channel are efficiently exploited.
More importantly, when two channel taps suddenly disappear in time slot $t=80$, the HB-Kalman algorithm requires about $20$ time slots to detect this, while the proposed D-OMP algorithm is able to detect the change immediately, which confirms the tracking capability of the proposed scheme as discussed in section IV-B.

\begin{figure}
\center{\includegraphics[width=0.5\textwidth]{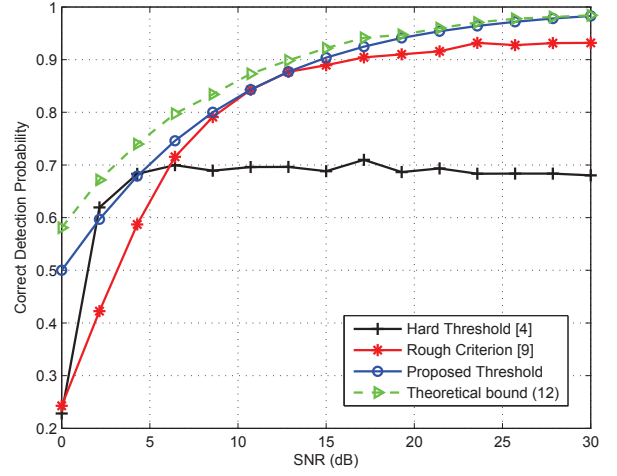}}
\vspace{-5mm}
\caption{The correct detection probability comparison against SNR.}
\vspace{-3mm}
\label{Detection_Probability}
\end{figure}

\begin{figure}
\center{\includegraphics[width=0.5\textwidth]{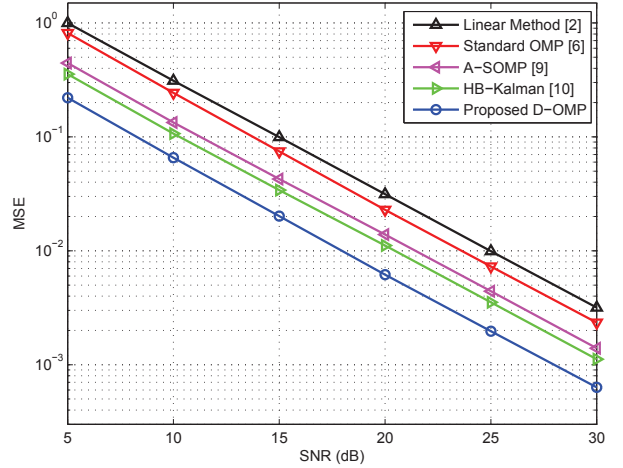}}
\vspace{-5mm}
\caption{MSE performance comparison against SNR.}
\vspace{-3mm}
\label{MSE_SNR_full}
\end{figure}

Fig. \ref{Detection_Probability} shows the correct detection probability of a persistent channel tap against SNR. It is evident that the hard threshold used in many CS-based channel estimation methods \cite{CSCE} is not adapted to the SNR. The rough criterion \cite{A-SOMP} can improve the correct detection probability by using the statistical information of the channel. For the proposed threshold $P_{th}$ based on noise statistics, the correct detection probability can be improved further, which is close to the theoretical bound derived in (\ref{theory_error}).

Fig. \ref{MSE_SNR_full} shows the MSE performance comparison against SNR for the five channel estimation methods.
It is clear that the standard OMP algorithm outperforms the linear method by about $1$ dB, where the benefit comes from utilizing the channel sparsity.
Further, A-SOMP and HB-Kalman are better than the standard OMP algorithm by about $2$ dB, since they partially consider the temporal correlations of the dynamic sparse channel.
For the proposed D-OMP algorithm, it is evident that another $2$ dB SNR gain can be achieved due to its capability to track the dynamic sparse channel rapidly and detect the non-zero channel taps accurately as discussed in Section IV.

\balance
\section{Conclusion}
In this paper, we have proposed a novel dynamic CS algorithm called D-OMP to rapidly track a dynamic sparse broadband communication channel in a sequential way by updating only the small channel variation over time.
%The proposed D-OMP algorithm merges the ideas from differential encoding and standard OMP algorithm, and adopt the threshold design based on noise statistics for the removal of disappearing taps.
An adaptive threshold based on noise statistics is used in the proposed D-OMP algorithm to remove the disappearing channel taps.
Moreover, the performance analysis provides the correct detection probability of persistent channel taps as well as the condition of appearing channel tap detection for the dynamic sparse channel with low computational complexity.
Finally, simulation results demonstrate that the proposed D-OMP algorithm is able to rapidly detect appearing channel taps, and achieves about $2$ dB gain compared with the recently proposed dynamic channel tracking algorithms.

\section*{Acknowledgment}
This work was supported by National Key Basic Research Program of China (Grant No. 2013CB329203) and National Natural Science Foundation of China (Grant Nos. 61271266 and 61411130156).

\newpage

\end{document}